\documentclass[aps,pre,reprint,superscriptaddress,twocolumn,showpacs]{revtex4-1}
\usepackage{amsfonts}
\usepackage{amsmath}
\usepackage{multirow}
\usepackage{graphicx}

\newcommand{\be}{\begin{equation}}
\newcommand{\ee}{\end{equation}}

\usepackage{newfloat,algcompatible}
\usepackage[size=small]{caption}
\usepackage{etoolbox}

\AtEndEnvironment{algorithm}{\noindent\hrulefill\par\nobreak\vskip-8pt}

\DeclareFloatingEnvironment[
    fileext=loa,
    listname=List of Algorithms,
    name=ALGORITHM,
    placement=tbhp,
]{algorithm}
\DeclareCaptionFormat{algorithms}{\vskip-15pt\hrulefill\par#1#2#3\vskip-6pt\hrulefill}
\algblock[Input]{Input}{EndInput}
\algblockdefx[Input]{Input}{EndInput}%
    [1]{\textbf{Input} #1}%
    {}
      
\begin{document}

% Use the \preprint command to place your local institutional report
% number in the upper righthand corner of the title page in preprint mode.
% Multiple \preprint commands are allowed.
% Use the 'preprintnumbers' class option to override journal defaults
% to display numbers if necessary
%\preprint{}

%Title of paper
%\title{Introducing the Scaling of Human Contacts in Reaction-Diffusion Processes}

\title{From calls to communities: a model for time varying social networks}

\author{Guillaume Laurent}
\affiliation{Laboratoire de l'Informatique du Parall\'elisme, INRIA-UMR 5668, IXXI, ENS de Lyon, 69364 Lyon, France}
\affiliation{BECS, Aalto University School of Science, P.O. Box 12200, FI-00076, Finland}
\author{Jari Saram\"{a}ki}
\affiliation{BECS, Aalto University School of Science, P.O. Box 12200, FI-00076, Finland}

\author{M\'arton Karsai}
\email[Corresponding author: ]{marton.karsai@ens-lyon.fr}
\affiliation{Laboratoire de l'Informatique du Parall\'elisme, INRIA-UMR 5668, IXXI, ENS de Lyon, 69364 Lyon, France}

\date{\today}

\begin{abstract}
Social interactions vary in time and appear to be driven by intrinsic mechanisms, which in turn shape the emerging structure of the social network. Large-scale
empirical observations of social interaction structure have become possible only recently, and modelling their dynamics is an actual challenge. Here we propose a temporal network model which builds on the framework of \textit{activity-driven time-varying networks} with memory. The model also integrates 
key mechanisms that drive the formation of social ties --  \textit{social reinforcement}, \textit{focal closure} and \textit{cyclic closure}, which have been shown
to give rise to community structure and the global connectedness of the network. We compare the proposed model with a real-world time-varying network of mobile phone communication and show that they share several characteristics from heterogeneous degrees and weights to rich community structure. Further, the strong and weak ties that emerge from the model follow similar weight-topology correlations as real-world social networks, including the role of weak ties.
\end{abstract}

% insert suggested PACS numbers in braces on next line
\pacs{}
% insert suggested keywords - APS authors don't need to do this
%\keywords{}

%\maketitle must follow title, authors, abstract, \pacs, and \keywords
\maketitle

% \begin{figure}
% \includegraphics{}%
% \caption{\label{}}
% \end{figure}

\section{Introduction}

The emergent structure and dynamics of social interactions are consequences of individual and collective decision-making processes~\cite{RevModPhys.81.591}. 
Understanding the driving forces behind social network formation has been a great challenge that has attracted lots of attention during the last decades. Even though conventional experimental approaches (based on e.g.~surveys) have revealed  fundamental rules of social behaviour~\cite{Wasserman1994Social}, it has not been possible scale such approaches up to sizes required for observing features of large-scale social structures, let alone their detailed dynamics. However, technological advances have helped to overcome these limitations through detailed digital recording of social interactions~\cite{Lazer2009Computational,Vespignani2009Predicting}. Following these advancements, a lot of effort has been put on mapping social structures from digital traces, and on establishing methodology for the analysis of social networks connecting millions of individuals. These efforts have led to discover the small-world architecture~\cite{Watts2002Simple}, the heterogeneous network structures induced by preferential attachment mechanisms \cite{deSollaPrice1965Networks,Albert2001Statistical}, the different roles of strong and weak ties~\cite{Granovetter1973Strength,Onnela2007Structure} and their relationship to triangle formation~\cite{Rapoport1957Contribution} and the emergence of communities~\cite{Fortunato2010Community, Kumpula2007Emergence}.

The earliest data-driven studies of social networks were mainly focused on static network structure. However, it is evident that this approach both unnecessarily discards available information -- electronic interaction traces typically come with time stamps -- as well as hinders detailed understanding of the mechanisms shaping network structure. In the recent years, steps have been taken to come around these limitations, with increased focus on both network dynamics (changes in link structure, such as in \cite{Palla2007Quantifying}) and temporal network structure~\cite{Holme2012Temporal} (time-stamped interaction events between individuals giving rise to \emph{e.g.}~temporal motifs~\cite{Kovanen2013Temporal}). In particular, studies of temporal networks of human interaction have revealed tie dynamics with memory and reinforcement processes~\cite{Zhao2011Social,Karsai2014Time}, bursty dynamical patterns \cite{Barabasi2005Origin,Barabasi2010Bursts},  temporal homophily in communication motifs~\cite{Kovanen05112013} as well as different strategies of communication and tie maintenance~\cite{Miritello2013Limited}.

Modelling temporal networks and their co-evolution with processes unfolding on the networks is still one of the key challenges in the field. Empirical studies employing null models with shuffled interaction sequences \cite{Karsai2011Small,1742-5468-2012-03-P03005,Miritello2011Dynamical} have pointed out that bursty interactions together with features such as weight-topology correlations strongly influence the speed of any type of information diffusing on the temporal network \cite{Lambiotte2013Burstiness,Takaguchi2013Bursty,PhysRevE.83.036102,Karsai2011Small,Miritello2011Dynamical}. However, a limiting factor of this approach is that it only allows to remove existing temporal inhomogeneities and correlations from empirical sequences in order to understand their effects. Because of this, the null model approach does not provide ways to continuously adjust the level of inhomogeneities or temporal correlations. This limitation can be overcome by resorting to artificial models of temporal networks~\cite{Rocha2013Epidemics,Perra2012Activity} that allow to fine-tune the level of different features one by one. One promising direction is the \emph{activity-driven framework} \cite{Perra2012Activity} that assumes heterogeneous activity of agents. It provides  a general foundation, which can be extended by incorporating additional temporal and structural correlations. Studies of the \emph{activity-driven time-varying network model} have demonstrated that memory effects can explain the emergence of strong and weak ties in social networks \cite{Karsai2014Time}, while endo- and exogeneous tie formation processes may control the evolution of business alliance networks \cite{Tomasello2014Role}. The same model has been used to study how time-varying interactions influence the dynamics of random walkers \cite{PhysRevLett.109.238701,Starnini2013Topological}, rumour spreading \cite{Karsai2014Time}, and epidemic processes \cite{PhysRevLett.112.118702}.

In this paper, our aim is to consider further social mechanisms that would make models of time-varying social networks more realistic. We introduce a temporal network model with adjustable community structure and emergent weight-topological correlations via the extension of the activity-driven time-varying network model with three additional mechanisms. These are i) reinforcement processes to model memory-driven interaction dynamics of individuals \cite{Karsai2014Time}, ii) focal and cyclic closure  motivated by the work of Kumpula et. al. \cite{Kumpula2007Emergence,Kumpula2009Model} to capture patterns responsible for the emerging community structure, and iii) a node removal process.
Using this temporal network model we demonstrate the effect of the scalable community structure and social reinforcement on information spreading, which co-evolves with the time-varying interactions. After the introduction of the model, we discuss its temporal behaviour, and match the emerging network properties with empirical observations. We discuss the present higher order correlations and their effects on spreading processes, and finally we conclude our work and discuss possible future directions.

\section{Model}

Our model definition builds on the activity-driven time-varying network (ADN) model introduced by Perra \emph{et al.}~\cite{Perra2012Activity}. The original model takes $N$ nodes that are each assigned \textit{a priori} with an activity probability per unit time $a_i=\eta x_i$. Here $x_i$ is the activity potential of  node $i$, drawn from a suitable  distribution $F(x_i)$ ($x_i\in [ \epsilon,1]$ bounded by a minimum activity $\epsilon$), and $\eta$ is a time rescaling factor determining the average number of active nodes per unit time to be $\eta \langle x \rangle N$. At the beginning of each time step (time $t$), the model network $G_t=(V_t,E_t)$ is initially disconnected ($E_t=\emptyset$) and it evolves via the following steps: (i) Through random sequential update each node $i$ becomes active with probability $a_i\Delta t$ and connects to $m$ other randomly selected nodes in $G_t$. (ii) At the end of each step (time $t+\Delta t$), all created links are removed and the process moves on to the next iteration. Note that during a single step, a node can participate in temporal interactions by creating a link (if it is active), or by receiving link(s) from other active nodes. The activity potentials of humans are commonly observed to be heterogeneous \cite{Cattuto2010Dynamics,Chmiel2009Scaling,Onnela2007Analysis}. We approximate this with a power-law activity potential distribution $F(x_i)\sim x_i^{-\gamma}$ with exponent $\gamma=2.8$, based on empirical observations and earlier modelling work \cite{Karsai2014Time}. Furthermore, since we aim to model one-to-one interactions, typical for mobile telephone communication, we set $m=1$. In addition, without loss of generality, we fix the parameters $\eta=1$, $\epsilon=10^{-3}$ and $\Delta t=1$ \cite{Karsai2014Time}.

\subsection{Social mechanisms}

The first mechanism to be integrated with the activity-driven model is \emph{memory} in terms of frequently repeated interactions between a node and its peers who have been contacted before. This has been considered earlier by Karsai \emph{et al.}~\cite{Karsai2014Time} with \textit{reinforcement processes}, where a node remembers its connected neighbours, and depending on its degree $n$ at $t$, either creates a new link with probability $p(n)=c/(n+c)$, or interacts with a neighbour with probability $1-p(n)=n/(n+c)$ to reinforce an existing tie. The introduction of memory defines a non-Markovian dynamics where past interactions influence the actual decision of a node. The larger a node's egocentric network size $n$, the smaller the chance for it to create a new link in the actual iteration. This captures the realistic assumption that interactions in a social network are not random but favoured with actual friends whose number is strongly limited by cognitive capacities \cite{Dunbar1992Neocortex,Saramaki2014Persistence}. This decision-making mechanism naturally leads to the emergence of weak and strong ties, weight heterogeneities, and a more realistic degree distribution in the integrated network structure \cite{Karsai2014Time}. Note that in this formulation $c$ scales the strength of memory, which may capture the attitude being a \textit{social keeper} or \textit{social explorer} \cite{Miritello2013Limited}. Here, for simplicity, we set $c=1$ for each node and leave the exploration of the effect of varying memory strengths for future studies \cite{UbaldiTo}.

The second mechanism we consider is that of tie creation with \emph{closure processes}. Here, \emph{cyclic closure}, the tendency to form network triangles, shapes the social structure at mesoscopic scale, and is responsible for the emergence of communities \cite{PhysRevE.90.042806}. The mechanism of \emph{focal closure}, on the other hand, is independent of network structure and represents the formation of ties between individuals with shared attributes or interests.  
It is driven by the propensity to seek cognitive balance between connected egos~\cite{Heider1946Attitudes,Granovetter1973Strength} as suggested by earlier theories in sociology~\cite{Wolff2011Sociology, Rapoport1957Contribution}. An applicative definition of \textit{cyclic} and \emph{focal closure} in general is given by Kumpula \emph{et al.}~\cite{Kumpula2007Emergence,Kumpula2009Model}, who model cyclic closure as biased local search. To the contrary, \emph{focal closure} is modelled as an unbiased global random search.

Finally, the third ingredient of the model is the process of \emph{node removal}. It allows the model network to reach an equilibrium state where its overall structural characteristics become invariant of time.

\subsection{Model definition}

Here we introduce a time-varying network model, which integrates memory and reinforcement processes with closure mechanisms in the activity-driven model definition. Our ultimate goal is to provide a temporal network model with adjustable community structure and weight-topology correlations to understand their role in shaping the emergent network structure and co-evolving dynamical processes.

Our model takes $N$ initially disconnected nodes and updates them in random order at each iteration step. In one iteration step, a node may become active with probability $a_i=\eta x_i$ or be deleted with probability $p_d$. Note that to keep the network size constant, for each deleted node, we add a new node to the network in the next iteration step. While a node $i$ is alive, it remembers its already connected neighbours $j\in V_t^{i}$ and the  weight $w_{ij}^t$ of interactions with each of them. 

If the node $i$ becomes active, then, depending on its actual degree $n_i$, it can attempt to form a new link with probability $p(n_i)=1/\left(n_i+1\right)$, or otherwise interact via an existing link with probability ($1-p(n_i)$). In the latter case it randomly selects one of its neighbours $j$ with probability $p^w_{ij}=w_{ij}^t/\sum_{k\in V_t^i} w_{ik}^t$ weighted by the number of their interactions. The two nodes then interact and increase their link weight $w_{ij}^t$ by $\delta$ (\textit{reinforcement process}). 

On the other hand, if the node decides to form a new link, it may follow different scenarios. In all cases, the new tie will initially have unit weight $w^t=1$.
If the degree of the focal node is $0$, it randomly picks another node from the entire network $j$ (\textit{focal closure}) and forms a tie. Otherwise, it attempts to create a new link with the triadic closure mechanism. First, it
chooses one of its neighbours $j$ randomly with a weighted probability $p^w_{ij}$. If $j$ has no other neighbours than $i$, node $i$ looks for another random node to interact with (\textit{focal closure}) and forms a link. Otherwise, it looks for a random neighbour $k$ of $j$ ($i\neq k$) with a weighted probability $p^w_{jk}$. If $k$ is not an already existing neighbour of $i$ ($k\notin V_t^i$), the two nodes interact with probability $p_{\Delta}$, and close the triad by forming a link. Otherwise, with probability $1-p_{\Delta}$, node $i$ follows the \textit{focal closure} mechanism and instead forms a link with a randomly selected node (other than $j$ and $k$). Finally, if $k$ is already a neighbour of $i$, that is, $k\in V_t^i$, the two nodes interact and increase the weight of their existing link by $\delta$ (\textit{reinforcement process}). At the end of each iteration step, all nodes finish their active interactions but remember their weighted egocentric network. For a pseudocode version of the algorithm, see Appendix A.

\section{Model network analysis}

In addition to the activity-driven model parameters whose values are fixed, our model has three intensive parameters, $p_{\Delta}$, $p_d$, and $\delta$. By varying these parameters, one can simulate a rich variety of time-varying networks with several emergent structural properties and correlations. In the following, we explore how the properties of the emerging network structure depend on time and on the intensive parameters, and whether those properties match with those of an empirical temporal network of mobile phone communications. The dataset we use here contains $633,986,311$ time-stamped mobile call communication records of $6,243,322$ customers of a mobile operator (market share $\approx 20\%$) in a European country. Customers are represented as network nodes, connected via $16,783,865$ weighted mutual links with weights defined as the number of calls between customers who mutually called each other at least once during the observation period. The presented network properties of the mobile phone call (MPC) network were calculated from a static, aggregated representation of the social network structure obtained by integrating the temporal interactions over $6$ months.

In the following, model networks were generated via large-scale numerical simulations with $N=10,000$ nodes, and results were averaged over $100$ independent realisations (if not noted otherwise). For each realisation, we measured the network parameters by considering links that are actually present in the network, \emph{i.e.}~we disregarded links of removed nodes.

\subsection{Temporal features}

The introduced network model is inherently temporal and simulates time-varying interactions between individuals. To explore its overall temporal behaviour, we measured two general network properties as a function of time. The first property is the average degree defined as $\langle k \rangle(t)=N^{-1}\sum_i k_i(t)$, where the sum runs over all nodes $i$, and $k_i$ denotes the number of established connections of node $i$ at time $t$ (i.e. $k_i(t)=|V_t^i|$). Second, we measure the average local clustering coefficient $C(t)$, defined as the fraction of the real and possible numbers of triangles around a node given its number of links, averaged over the whole network \cite{Newman2010Networks}. $C$ quantifies the density of triangles in a network; in social networks, the existence of communities typically gives rise to high triangle densities.

\begin{figure}[h!]
\centering
\includegraphics[width=1.0\linewidth]{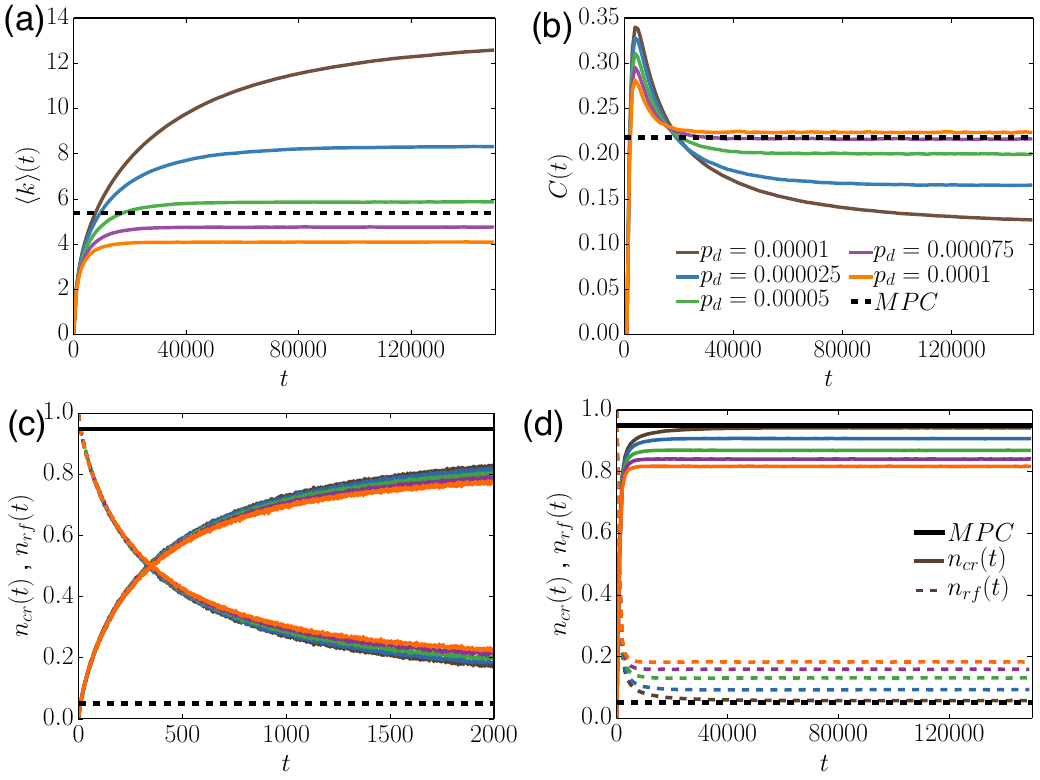}
 \caption{\label{fig:2} Average network properties as the function of time. We depict the evolution of a) the average degree $\langle k \rangle(t)$, b) clustering coefficient $C$, and c,d) the fraction of events creating ($n_{cr}(t)$) or reinforcing ($n_{rf}(t)$) a link with various values of the deletion probability  $p_d$ (for exact values, see legend). For each calculation, we fixed $p_{\Delta}=0.5$ and $\delta=1$.}
\end{figure}
 
\begin{figure*}[t!]
\centering
\includegraphics{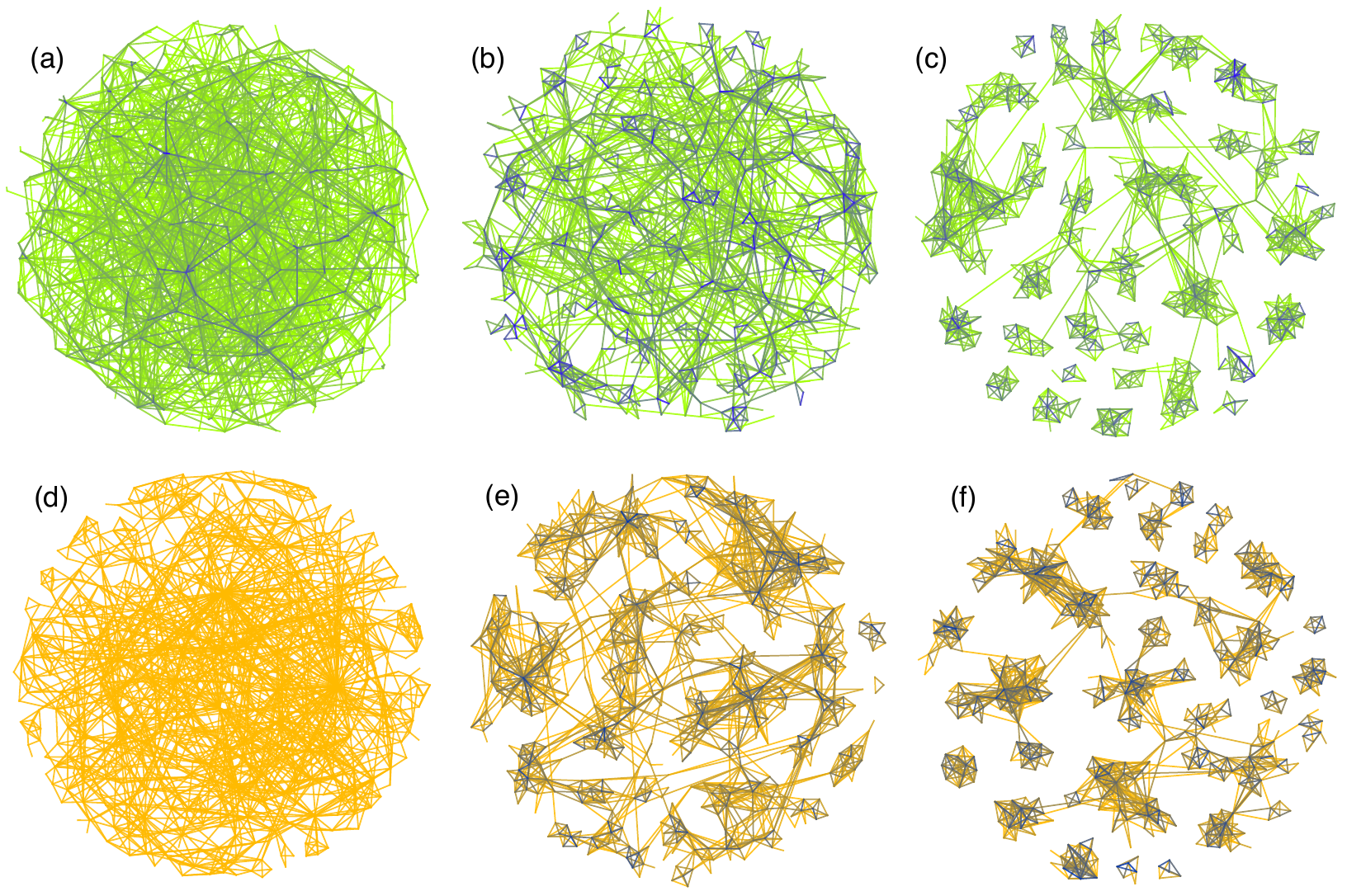}
 \caption{\label{fig:1} Demonstration of the emerging structure in the time-varying network model. Panels (a-c) depict simulated networks with fixed $\delta=1$ and varying $p_{\Delta}=0.5$, $0.9$, and $0.995$ ($p_d=4e-5$, $2e-5$, and $1.04e-5$) respectively. Panels (d-f) shows networks with fixed $p_{\Delta}=0.995$ with varying $\delta=0$, $0.5$, and $1.5$ ($p_d=3.5e-5$, $1.7e-5$, and $8e-6$) respectively. Each panel depicts the actual structure of a network with $N=500$, in its stationary state. Links are coloured according to their weight (darker link colour = stronger link weight).}
\end{figure*}

As follows from the model definition, the process starts from a set of disconnected agents, thus the measured properties are trivially zero at time $t=0$. However, as time goes by and ties are formed via temporal interactions, $\langle k \rangle(t)$ starts increasing until the network reaches a stationary state with constant average degree. The time it takes to reach this equilibrium state strongly depends on the choice of the node deletion probability  $p_d$  as shown in Fig.\ref{fig:2}.a. If  $p_d$  is too small, nodes remain  in the system for a long time and even nodes with small activity levels have time to evolve their egocentric network resulting in a slow relaxation to the stationary state. On the other hand, as $p_d$ is increased due to the finite life-time of nodes, less active nodes are removed before their egocentric structures are fully evolved. Because of this, the network reaches equilibrium faster. This explains the decrease of stationary average degree with increasing  $p_d$. The average degree can be tuned approximate the empirical average degree of the MPC network (dashed line in Fig.\ref{fig:2}.a) 
 
Measurement of the time dependence of the  the average clustering coefficient (see Fig.\ref{fig:2}.b) yields a fairly similar scenario. However, here,  after the initial increase of $C(t)$ it reaches a maximum,  followed by a decrease and relaxation to a stationary value. This is because triangles tend to emerge right in the beginning of the process,  followed by the evolution of strong ties. After nodes have created their first links and closed triangles between them, interactions begin to frequently take place on existing links that are reinforced in the process. Thus the local search of active nodes is biased towards strong ties that are part of already formed triangles. In late times nodes attempt less frequently to create new triangles through the weak links that emerge throughout the dynamics by focal closure. Again, there is saturation due to node removal influenced by the rarely active but surviving nodes, who keep introducing random links in the network. By choosing an appropriate parameter values, the clustering coefficient can again be tuned to  comparable to its corresponding empirical value (dashed line in Fig.\ref{fig:2}.b). 

The above picture is fully supported by measurements capturing the fraction of events which create new links $n_{cr}(t)=\#e_{cr}/(\#e_{cr}(t)+\#e_{rf}(t)$), or reinforce already existing ones $n_{rf}(t)=\#e_{rf}/(\#e_{cr}(t)+\#e_{rf}(t)$). Here $\#e_{cr}$ ($\#e_{fr}$) denotes the number of such events at time $t$. As seen in Fig.\ref{fig:2}.c these measures take values of $1$ and $0$ (respectively) at time $t=0$, as all events initially create new links. However they rapidly approach a constant value as it is visible in Fig.\ref{fig:2}.d. On the other hand, as $p_d$ is increased, the number of events reinforcing already existing links is increased, while events responsible for the creation of new links become less frequent. This is in accordance with our  interpretation above. The measures can be tuned to reflect empirical network values here as well (black lines). 

In the following, if not noted otherwise, we set $p_d=0.5\times 10^{-5}$ and run the simulations for $t=150,000$ time steps to ensure the model networks to always reach their stationary state with average properties invariant in time.

\subsection{Static features}
Next we concentrate on the effect of different mechanisms on the emerging network structure in the stationary state. In Fig.\ref{fig:1}, we visually demonstrate the role of the cyclic closure (a-c) and link weight reinforcement (d-f) mechanisms in the emerging model networks. In panels Fig.\ref{fig:1}(a-c), we have kept the reinforcement increment constant, $\delta=1$, while varying the cyclic closure probability $p_{\Delta}$ between $0.5$ and $0.995$. The node deletion probability $p_d$ has been chosen to yield networks with suitable link density for visualisation (for exact parameters see the figure caption). When $p_{\Delta}$ is small, the emerging network structure is densely connected and appears more like a random structure since link creation is driven by focal closure. Nevertheless, because of the reinforcement process, weight heterogeneities emerge already in this case. More interestingly, by increasing $p_{\Delta}$, communities are seen to emerge, with weight-topology correlations that are in line with the Granovetterian picture \cite{Granovetter1973Strength,Onnela2007Structure}, where strong ties connect nodes inside communities, while weak ties  emerge between them (darker link colour = stronger link weight). However, focal closure alone is not sufficient for the emergence of community structure as shown in panels Fig.\ref{fig:1}(d-f). Here, the triangle formation probability has been kept constant, $p_{\Delta}=0.995$, while the reinforcement parameter  $\delta$ has been varied between $0$ and $1.5$. Even though a triangle would almost always be closed by the local search if a suitable node were found, without link weight reinforcement (d), the local search will be disperse and often resulting in focal closure because hitting a neighbour with no other neighbours. However, by increasing $\delta$ the local search will get more biased towards emerging strong ties and lead to the evolution of tight community structure. Note that since $\delta$ controls the strength of the local search mechanism, it also scales the size of the communities.

\begin{figure}[h!]
\centering
\includegraphics[width=1.0\linewidth]{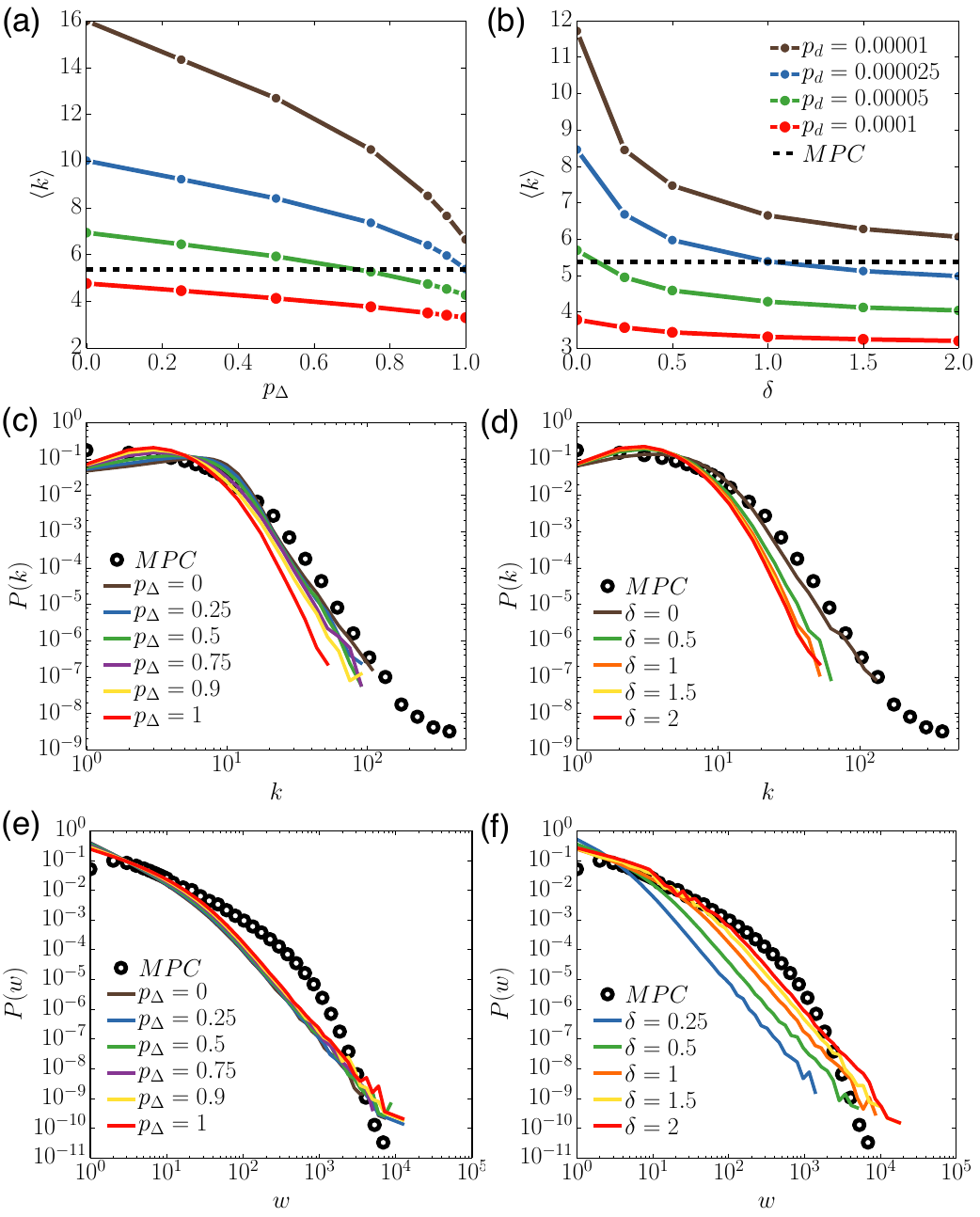}
 \caption{\label{fig:3} General network measures as a function of the intensive parameters. (a) and (b) show the dependence of the average degree $\langle k \rangle$ on $p_\Delta$ and $\delta$, respectively (for the former, $\delta=1$ and the latter, $p_{\Delta}=1$). Various values of  $p_d$ have been used (see legend in (b)). The dashed black line depicts the average degree $\langle k \rangle$ of the empirical MPC network. (c) and (d) depicts the degree distributions  $P(k)$ of model networks with varying $p_{\Delta}$ and $\delta$, while (e) and (f) show the corresponding weight distributions $P(w)$ . In panels (c) and (e) ((d) and (f)), we kept $\delta=1$ ($p_{\Delta}=1$) and $p_d=5\times 10^{-5}$. Black circles denote the corresponding MPC distributions.}
\end{figure}

To quantitatively explore the emerging network structure as the function of the intensive parameters, we have measured the behaviour of general network properties. Fig.\ref{fig:3} shows how the average degree of the model networks depends on $p_{\Delta}$  and $\delta$ (panels a and b, respectively) for different deletion probabilities  $p_d$.  The average degree of the empirical MPC network is displayed for comparison (dashed horizontal line). In each case, the average degree decreases when the parameters controlling the probability of cyclic closure $p_{\Delta}$ and the amount of link reinforcement $\delta$ are increased, indicating that these parameters together with $p_d$ strongly control $\langle k \rangle$. However, while the dependence of $\langle k \rangle$ on $p_{\Delta}$  is concave,  $\langle k \rangle (\delta)$ is a convex function for any $p_d$. Moving beyond averages, 
panels c) and d) display the behaviour of the degree distributions $P(k)$ for various $p_\Delta$ and $\delta$, indicating that there are strong degree heterogeneities. While the average degrees strongly depend on these parameters, the overall distribution shapes do not change much. The distribution of the empirical MPC is again displayed for reference, showing a very similar shape than those produced by the model.

The strength of social ties, measured here as the number of dyadic interactions, is commonly observed to be distributed heterogeneously in social networks. This is also the case with our empirical MPC network (black circles in Fig.\ref{fig:3} e and f). Tie strength heterogeneity is an emergent property of our model networks where it comes from the preference of nodes to reinforce existing links. This mechanism is independent from the search strategies, as evident in Fig.\ref{fig:3}.e, where the model weight distributions are invariant to the selection of the cyclic closure probability $p_{\Delta}$  ($\delta$ and $p_d$ are kept constant here). The emerging weight heterogeneities are the consequence of the intrinsic memory process, which inclines nodes to reinforce existing links rather than create new ones. Therefore, the weight heterogeneity naturally depends on the reinforcement parameter $\delta$ (see Fig.\ref{fig:3}.f), which in turn mainly scales  the tail of $P(w)$ but does not affect not its functional shape, which appears to be a power-law with a constant exponent.

\subsection{Higher-order correlations}

Other than the observed degree and weight heterogeneities in the model networks, certain parameter ranges result in the emergence of rich community structure. In real social networks, communities can be characterised by higher-order correlations as they are built of  closed triangles with unevenly distributed links of various strength in the structure. Stronger ties that are maintained by frequent interactions tend to connect nodes inside communities and shape the structure locally, while weak links with infrequent interactions are situated between communities and keep the social structure globally connected. To show whether the modelled temporal networks can recover these characteristics, we have performed three sets of measures and compared our findings to the empirical results.

\begin{figure}[h!]
\centering
\includegraphics[width=1.0\linewidth]{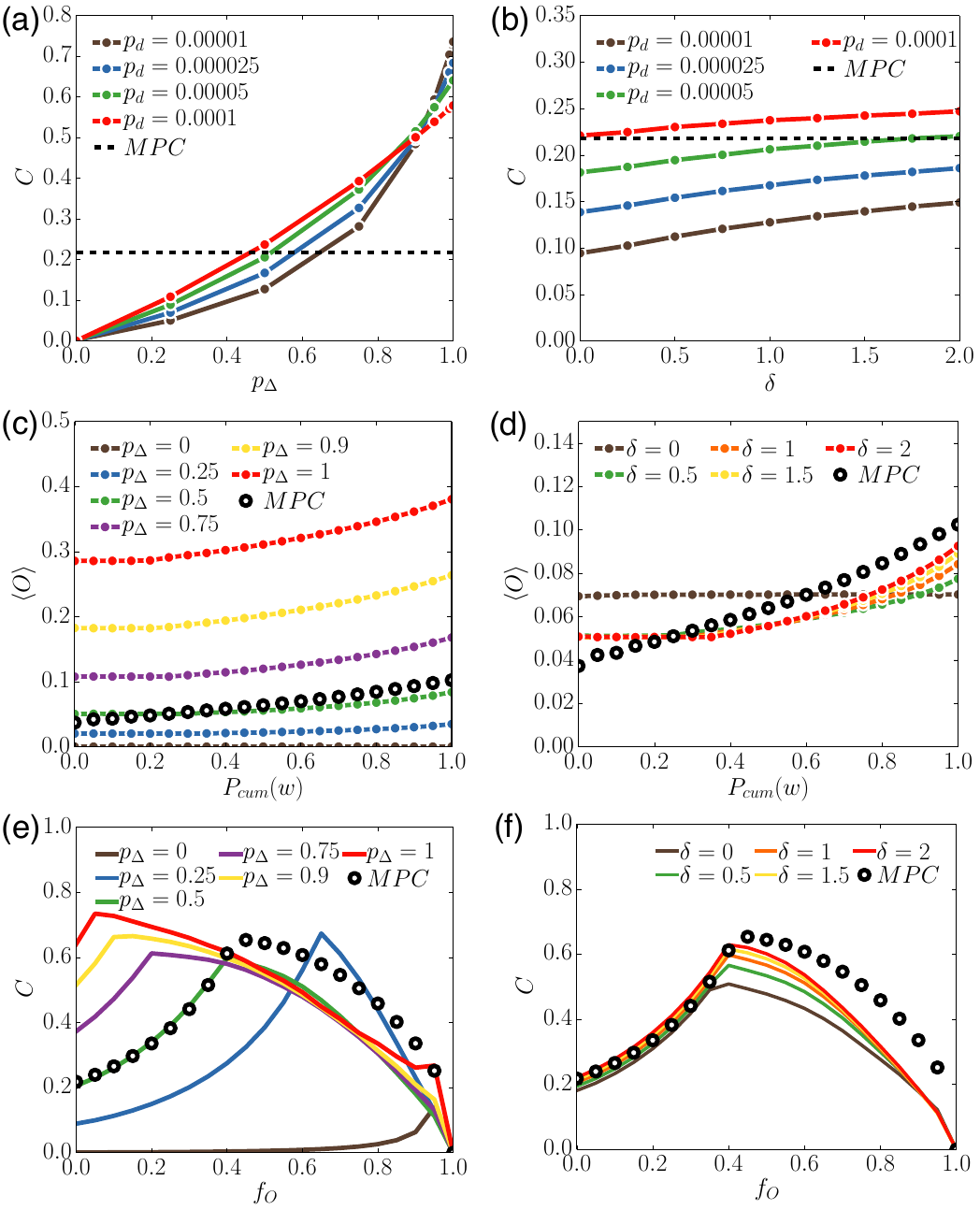}
 \caption{\label{fig:4} Measures of higher-order correlations in the empirical and model networks. Panels (a) and (b) show the dependence of the average local clustering coefficient $C$ on the parameters $p_{\Delta}$ and $\delta$, respectively, for the empirical and model networks. Weight-topology correlations are shown as the function of the same parameters ((c) and (d), respectively) by measuring the average overlap $\langle O \rangle$ as the function of cumulative tie strength $P_{sum}(w)$. Panels (e) and (f) depict the average local clustering coefficient of the residual networks after removing a fraction $f_O$ of nodes ranked by their overlap. Empirical results are shown in each panel with black dashed lines or circles, while measures on model networks are depicted by coloured solid lines with the corresponding parameters in the legend. In each measurements in panels (a,c,e) we kept $\delta=1.0$, in panels (b,d,f) $p_{\Delta}=0.5$, while in panels (c-f) $p_d=5\times 10^{-5}$.}
\end{figure}

As we have seen earlier, both the values of $p_{\Delta}$ and $\delta$ influence the community structure. However, we expect that $p_{\Delta}$ plays a dominant role here as it controls the triangular closing mechanism in the network. If $p_{\Delta}=0$, links are created randomly and the clustering coefficient is very small as seen in Fig.\ref{fig:4}.a, where a constant $\delta=1$ were set for each measurement. By increasing $p_{\Delta}$, cyclic closure becomes more dominant, reflected in an increasing $C$ with very high values as $p_{\Delta}$ goes to its maximum. The clustering coefficient of the MPC network is depicted as the horizontal dashed line. There are several sets of parameter values for which the model yields networks with similar clustering coefficient values, \emph{e.g.}~$p_{\Delta}\simeq 0.5$, $\delta\simeq 1$, and $p_d\simeq 5\times 10^{-5}$ (see the crossing points of the black dashed and green solid lines in Fig.\ref{fig:4}.a). Not surprisingly, $\delta$ plays a weaker role in the emergence of triangles. The reinforcement mechanism introduces a bias in the local search for new ties, which leads to tighter communities and more closed triangles as we increase $\delta$, reflected by the slightly increasing $C$ in Fig.\ref{fig:4}.b.

Second, we have measured weight-topology correlations to check if the model networks are constructed according to the Granovetterian weak-tie structure. In his seminal paper \cite{Granovetter1973Strength} Granovetter suggests that the fraction of common friends of connected individuals is positively correlated with their tie strength (\emph{i.e.}~their link weight here). The fraction of common friends can be measured by the link overlap \cite{Onnela2007Structure} defined as $O_{ij}=n_{ij}/((k_i-1)+(k_j-1)-n_{ij})$ where $n_{ij}$ is the number of common neighbours of nodes $i$ and $j$, and $k_i$ and $k_j$ are their degrees. To quantify weight-topology correlations, we measure  the average link overlap $\langle O \rangle$ as a function of cumulative tie strength $P_{cum}(w)$ that measures the fraction of links with tie strength smaller than $w$ \cite{Onnela2007Structure}. In the MPC network, this function reflects positive correlations between overlap and tie strength (black circles in Fig.\ref{fig:4}.c and d) in accordance with earlier observations \cite{Onnela2007Structure}. If the model networks have their triangle-formation mechanism or weight reinforcement turned off ($p_{\Delta}=0$ or $\delta=0$), no such correlation is found. However, for any positive values of $p_{\Delta}$ and $\delta$, a positive weight-topology correlation emerges in accordance with the empirical observation and the hypothesis of Granovetter. For larger $p_{\Delta}$ (with constant $\delta=1$), the function is shifted as more triangles evolve, which is increasing the average overlap (see Fig.\ref{fig:4}.c), while by increasing $\delta$ (and constant $p_{\Delta}=0.5$) (see Fig.\ref{fig:4}.d) the correlation becomes stronger as ties with larger strength appear in the networks.

At last, we have checked how the clustering coefficient $C$ changes by removing a fraction of links $f_O$ in increasing order of overlap. The resulting functionf $C(f_O)$ can be divided into two regimes. Trivially the removal of links with $O=0$, which connect communities, does not decrease the number of triangles in the network but only decreases the degrees of nodes by removing links not participating in triangles. Consequently, by the removal of these links the clustering coefficient must increase. On the other hand, the removal of links with $O>0$ that connect nodes inside communities reduces the number of triangles, resulting a decrease in $C$. This behaviour is observed in the empirical system (black circles in Fig.\ref{fig:4}.e and f) and also for the model networks. In Fig.\ref{fig:4}.e as $p_{\Delta}\rightarrow 0$ no triangles evolve in the network, and thus most of the links have $O=0$ , and the first regime is extended. In the other extreme, as $p_{\Delta}\rightarrow 1$, most of the links have non-zero overlap and the second regime dominates. The best match with the empirical curve appears when $p_{\Delta}\simeq 0.5$ (red curve in Fig.\ref{fig:4}.e) and $\sim40\%$ of links evolve with $O=0$. Note that the second regime of this function is very sensitive to the fine-grained structure of the actual communities, which causes the difference between the model and empirical curves. In addition, by keeping $p_{\Delta}=0.5$, we have checked the $\delta$ dependence of this function (see Fig.\ref{fig:4}.f). As the  tie reinforcement $\delta$ increases, links form tighter communities and the clustering coefficient increases in this case.

\section{Effects on spreading processes}

\begin{figure}[h!]
\centering
\includegraphics[width=1.0\linewidth]{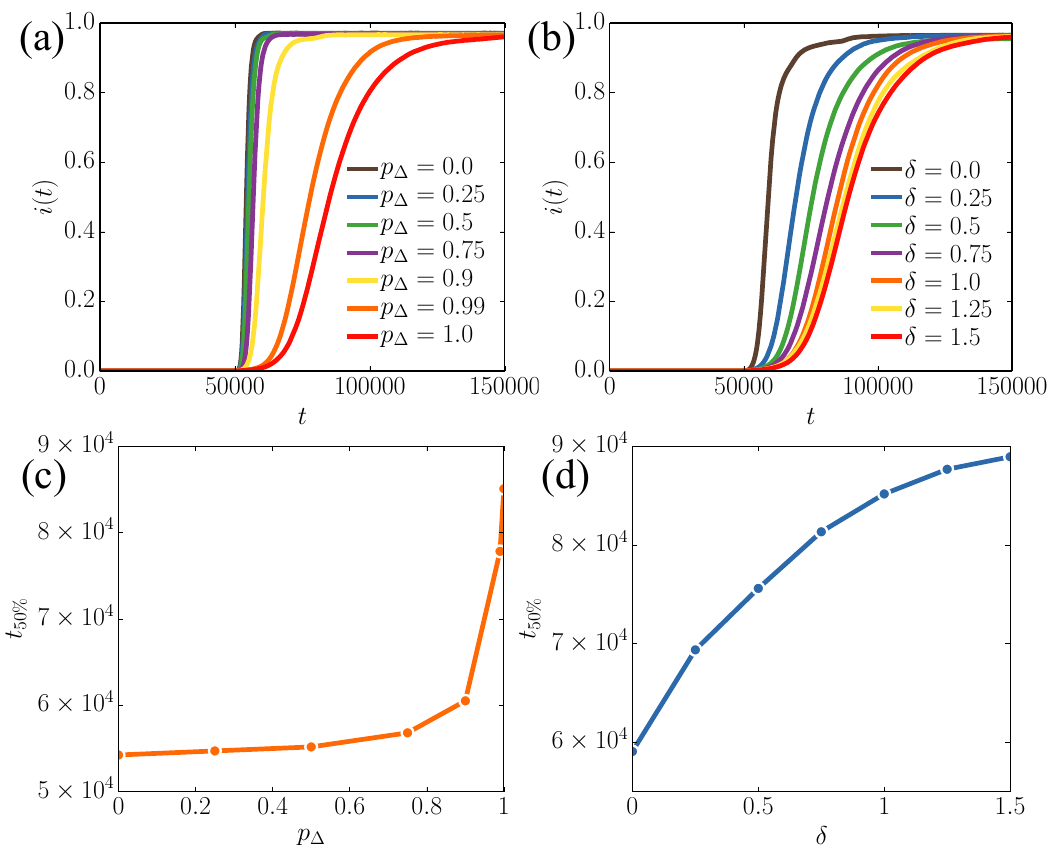}
 \caption{\label{fig:5} Susceptible-Infected process co-evolving with the time-varying network model (a, b) Fraction of infected nodes as a function of time for different values of $p_{\Delta}$ and $\delta$  (for exact parameters see legend). (c, d) Time $t_{50\%}$ of $50\%$ penetration for different values of $p_{\Delta}$ and $\delta$. While in case of panels (a, c) we have kept $\delta=1$,  for panels (b, d) we have used  $p_{\Delta}=0.99$. Each simulation was run for $t=150,000$ iteration steps with $p_d=t\times 10^{-5}$ on networks of size $N=10,000$. Results were averaged over $100$ realisations, where we initiated the process from a single random seed at $t=50,000$.}
\end{figure}

We conclude our modelling study by demonstrating its capacity in simulating dynamical processes co-evolving with the temporal network structure with scalable integrated social mechanisms. For this reason, we have selected the simplest possible spreading process, the susceptible-infected (SI) model \cite{Barrat2008Dynamical}, where each node can be in one of two mutually exclusive states; susceptible ($S$) or infected ($I$). Initially, every node is in the state $S$, except for a random seed with state $I$. The infection passes from any node in the $I$ state to any node in the $S$ state via a temporal  interaction, but independent of the direction of the actual temporal link \cite{Karsai2011Small}. In order to assure that the network is in the stationary state we initiated the spreading process after the temporal network of size $N=10,000$ has evolved for $t=50,000$ iterations. After this time we measured the $i(t)=I(t)/N$ fraction of infected nodes until $t=150,000$, where the process has reached maximal penetration (see Fig.\ref{fig:5}.a and b). Note that maximal penetration is not  converging to $1$ as there are always new susceptible nodes introduced in the network, moreover the temporal structure is not necessarily connected. For simplicity, we set the probability of transmission per interaction event to $\lambda=1$ (note that then the speed of transmission between two nodes is only limited by the frequency of their interactions). Further, we consider interactions as undirected and therefore they may transmit the interaction both ways. 

The simulation results depicted in Fig.\ref{fig:5}.a and b show that both the spreading process becomes slows when either the cyclic closure probability $p_{\Delta}$  or the weight reinforcement parameter $\delta$ are increased. While in the first case this is mainly due to the emerging tight communities, in the second case weight-topology correlations also play a role. The same conclusions can be drawn by measuring the speed of spreading, quantified as the time $t_{50\%}$ when the process has reached $50\%$ penetration. Tight communities evolve as $p_{\Delta}\rightarrow 1$, which constrains the spreading process and explains the convex shape of the curve in Fig.\ref{fig:5}.c. On the other hand, weight-topology correlations appear even for small values of $\delta$ (as we have shown in Fig.\ref{fig:4}.c). These immediately slow down the speed of spreading and cause the concave shape of the function in Fig.\ref{fig:5}.d. Thus, one can conclude that even though more pronounced community structure has considerable effects on the spreading process, already weak weight-topology correlations can strongly affects the speed of spreading.

\section{Conclusions}

The temporal network representation takes into account the time-varying nature of interactions between entities instead of considering only the static network structure of their connections \cite{Holme2012Temporal}. So far, this approach has mainly focussed on analyzing empirical data, while there is still a lack of appropriate network models. In this paper we introduce a model of temporal social networks with emerging heterogenous structure, communities, and higher-order correlations. As a starting point, we take the recently introduced modelling framework of activity-driven networks with heterogeneously distributed activities of individuals \cite{Perra2012Activity} and memory processes \cite{Karsai2014Time}. We extend this model with three mechanisms that have been earlier used for generating static weighted networks~\cite{Kumpula2007Emergence}, such as \textit{social reinforcement}, \textit{cyclic} and \textit{focal closure}. These mechanisms arguably drive the tie formation of individuals and are responsible for the emergence of communities and the global connectedness of the network. In addition, we introduce a \textit{node removal} process allowing the model network to reach an stationary state, where its overall structural characteristics become invariant of time. By adjusting the model parameters, it is possible to recover several characteristics of a real-world temporal network of mobile phone calls. These include heterogeneous degrees and weights, rich community structure, and weight-topology correlations and weak tie properties as suggested by Granovetter \cite{Granovetter1973Strength}.

The main advantages of this model are that (a) it is able to mimic the time-varying nature of interactions characterising several real systems; (b) it helps to understand the importance of different mechanisms in shaping the emerging network structure; and (c) it allows us to test their effect in a scalable fashion on co-evolving dynamical processes. This last point is of special importance, as \emph{e.g.}~spreading processes have been shown to have significantly different critical behaviour when co-evolving with time varying interactions. In addition, by varying the model parameters one can control the average degree, strength of ties, size of communities, clustering, and interconnectedness of the network. These features can control e.g. the speed of spreading as we demonstrated via a simple study on SI processes, but play crucial effects in case of other processes like random walks, the diffusion of information, epidemics, or social contagion processes, making the model an ideal testbed for simulating such processes.

% If you have acknowledgments, this puts in the proper section head.
\begin{acknowledgments}
JS acknowledges support by the Academy of Finland,  project  no.   260427 ("Temporal Networks of Human Interactions"). We thank A.-L.~Barab\'{a}si for providing the MPC data.
\end{acknowledgments}

\appendix

\section{Algorithmic model definition}

Here we describe the algorithmic implementation of the temporal network model introduced in the main text. The main function \textit{\textbf{Temporal network model}($G_{t=0}$, $T$, $p_{\Delta}$, $p_d$, $\delta$)} takes as input a weighted network with empty link set $G_{t=0}=(V,\emptyset,\{ a_i\}, \{w^t_{ij}\})$ with size $N$,  pre-assigned node activity probabilities per unit time $\{ a_i\}$, and $ initial link weights \{w^{t=0}_{ij}=0\}$ . Further parameters are $T$: the number of iterations; $p_{\Delta}$: probability of triadic closure; $p_d$: probability of node deletion; and $\delta$: link reinforcement increment. In the pseudo code, $n_i=|V_{t}^i|$ denotes the current degree of node $i$; $rand()$ is a pseudo-random number generator, which returns a rational number between $[0,1]$; and $p^w_{ij}=w_{ij}^t/\sum_{k\in V_{t}^i} w_{ij}^t$ is the probability to select a random neighbour $j$ from the current neighbour set $V_t^i$ of node $i$ weighted by the number of interactions between them performed since the link was created up to time $t$ ($\sum_{j\in V_t^i}p^w_{ij}=1$). Temporal events between nodes $i$ and $j$ performed at the current iteration step $t$ are denoted by $[t,i,j]$.
\begin{algorithm}[]
\caption {\textit{\textbf{Temporal network model}}}
\begin{algorithmic}
\Input{: $G_{t=0}$, $T$, $p_{\Delta}$, $p_d$, $\delta$}\EndInput
\FOR{$t=1$ to $T$}
\FOR{$N$ randomly selected node $i \in V$}
	\IF{$rand()\leq p_{d}$}
		\STATE{remove all links of node $i$}
		\COMMENT{\textit{delete and re-insert $i$}}
	\ENDIF
	\IF{$rand()\leq a_i$}
	\COMMENT{\textit{node $i$ is active}}
		\IF{$rand() \leq p(n_i)$}
		\COMMENT{\textit{link creation}}
			\IF{$n_i=0$}
			\COMMENT{\textit{node $i$ has no neighbour}}
				\STATE{\textit{\textbf{Focal closure}($i$, $\{ i \}$, $G_{t}$)}}
			\ELSE
				\STATE{\textit{\textbf{Cyclic closure}($i$, $G_t$)}}
			\ENDIF
		\ELSE
		\COMMENT{\textit{link reinforcement}}
			\STATE{Select a $j$ neighbour of $i$ with probability $p^w_{ij}$}
			\STATE{Event $[t,i,j]$}
			\STATE{$w_{ij}^t+=\delta$}
		\ENDIF
	\ENDIF

\ENDFOR
\ENDFOR
\end{algorithmic}
\end{algorithm}

The main function calls two subroutines called \textit{\textbf{Cyclic closure}($i$, $G_t$)} and \textit{\textbf{Focal closure}($i$, $G_t$, $X$)} with input parameters $i$ denoting the currently active node; $G_t$ the current weighted network structure; and $X$ exception set of nodes to connect in the current call.

\begin{algorithm}[]
\caption {\textit{\textbf{Cyclic closure}}: as input $i$ is the current active node; and $G_t$ is the current weighted network}
\begin{algorithmic}
\Input{: $i$, $G_t$}\EndInput
\STATE{Select a $j$ neighbour of $i$ with probability $p^w_{ij}$}
\IF{$n_j=1$}
	\COMMENT{\textit{node $j$ has one neighbour}}
	\STATE{\textit{\textbf{Focal closure}($j$, $\{ i,j \}\cup V_t^{i}$, $G_{t}$)}}
\ELSE

	\STATE{Select a $k(\neq i)$ neighbour of $j$ with probability $p^w_{jk}$}
	\IF{$(i,k)\notin E_t$}
	\COMMENT{\textit{triangle $(i,j,k)$ is not closed}}
		\IF{$rand() < p_{\Delta}$}
			\STATE{Event $[t,i,k]$}
			\STATE{$w_{ik}^t=1$}
		\ELSE
			\STATE{\textit{\textbf{Focal closure}($j$, $\{ i,j,k \}\cup V_t^{i}$, $G_{t}$)}}
		\ENDIF
	\ELSE
		\STATE{Event $[t,i,k]$}
		\STATE{$w_{ik}^t+=\delta$}
	\ENDIF
\ENDIF
\end{algorithmic}
\end{algorithm}

\pagebreak

\begin{algorithm}[]
\caption {\textit{\textbf{Focal closure}}: as input $i$ is the current active node; $X$ is the exception set of nodes to connect; and $G_t$ is the current weighted network}
\begin{algorithmic}
\Input{: $i$, $X$, $G_{t}$}
\EndInput
\STATE{Select a random node $j \in V \setminus X$}
\STATE{Event $[t,i,j]$}
\STATE{$w_{ij}^t=1$}
\end{algorithmic}
\end{algorithm}

% Create the reference section using BibTeX:

\end{document}